\documentclass[aps,twocolumn,floatfix,amsmath,amssymb]{revtex4-1}

\usepackage{amsmath,bm,amsfonts}
\usepackage{graphicx}

\newfont{\bg}{cmr10 scaled\magstep4}

\newcommand{\bigzerou}{\smash{\lower1.7ex\hbox{\bg 0}}}

\begin{document}

\title{Closest Wannier functions to a given set of localized orbitals}
  
\author{Taisuke Ozaki}
\affiliation{
  Institute for Solid State Physics, The University of Tokyo, Kashiwa 277-8581, Japan
}

\date{\today}

\begin{abstract} 

A non-iterative method is presented to calculate the closest Wannier functions (CWFs) to a given set of localized guiding functions,
such as atomic orbitals, hybrid atomic orbitals, and molecular orbitals, based on minimization of a distance measure function.   
It is shown that the minimization is directly achieved by a polar decomposition of a projection matrix via singular value decomposition, 
making iterative calculations and complications arising from the choice of the gauge irrelevant.
The disentanglement of bands is inherently addressed by introducing a smoothly varying window function and 
a greater number of Bloch functions, even for isolated bands. 
In addition to atomic and hybrid atomic orbitals, we introduce embedded molecular orbitals in molecules and bulks 
as the guiding functions, and demonstrate that the Wannier interpolated bands accurately reproduce the targeted conventional bands 
of a wide variety of systems including Si, Cu, the TTF-TCNQ molecular crystal, and a topological insulator of Bi$_2$Se$_3$. 
We further show the usefulness of the proposed method in calculating effective atomic charges. 
These numerical results not only establish our proposed method as an efficient alternative for calculating WFs, 
but also suggest that the concept of CWF can serve as a foundation for developing novel methods to analyze electronic 
structures and calculate physical properties.

\end{abstract}

\maketitle

\section{INTRODUCTION}

The Wannier functions (WFs) \cite{Wannier37,Kohn59,Marzari12} play a central role in analyzing electronic structures 
of real materials and advancing electronic structure methods alongside density functional theory (DFT) \cite{Hohenberg1964} 
and the other electronic structure theories \cite{Hamann09,Lechermann06}.
A widely adopted method for calculating WFs involves maximizing their localization, which can be reformulated as minimizing 
the spread function characterizing the variance of WFs in real space \cite{Marzari97}. 
The concept of maximal localization leads to an elegant formulation, resulting in maximally localized Wannier functions (MLWFs),  
and the compact representation of the Hamiltonian that follows allows us to analyze the electronic structures of real materials 
using a localized orbital picture. 
However, it is also recognized that the minimization of the spread function can often encounter local minima, 
particularly in large-scale systems having complicated electronic structures  \cite{Marzari12}. 
To overcome this challenge, methods aimed at automated high-throughput Wannierisation have been developed \cite{Vitale20,Damle18,Qiao23}.
In addition, the other methods for generating (nonorthogonal) WFs have been proposed based on projection methods \cite{Ku02,Lu2004,Qian08}, 
which produce compact atomic like orbitals. The orthogonalization of such nonorthogonal orbitals obtained by the projection can be 
achieved by the L\"owdin orthogonalization procedure \cite{Lowdin1950,Cloizeaux1964}. 
These orthogonalized orbitals are utilized as an initial guess of 
WFs in the minimization of the spread function to obtain MLWFs \cite{Marzari97,Mostofi2008}. 
Apart from the role of initial guess, it is worth noting that the L\"owdin orthogonalized orbitals possess an important variational property, 
which in particular minimizes the sum of squared distances between the orthogonal and original nonorthogonal orbitals 
in the Hilbert space \cite{Carlson57}. 
As long as the original nonorthogonal orbitals are localized, the localization of the L\"owdin orthogonalized 
orbitals in real space is guaranteed in a sense of minimization of the distance. 
In quantum chemistry, this property is leveraged to develop methods for generating localized orthogonal orbitals. 
These methods, which include the natural atomic and bond orbital method \cite{Reed88} 
and the intrinsic atomic orbital method \cite{Knizia13}, involve obtaining orthogonalized orbitals through 
L\"owdin type orthogonalization, giving heavy weight to occupied states via the density matrix.
However, in solid state physics the variational property inherent in the L\"owdin orthogonalization 
has not been fully explored yet. A reason for this is that the L\"owdin orthogonalization requires calculation of 
$S^{-1/2}$ for the overlap matrix $S$, which causes numerical instability for nearly ill-conditioned matrices. 
Due to the difficulty, advancements in methodologies along these lines appear to be slow.
Our study aims to develop an efficient and robust method to generate the closest Wannier functions (CWFs) to a given set 
of localized guiding functions. We will demonstrate that exploiting the variational property leads to a versatile method 
that eliminates the need for iterative optimization and avoids complications related to gauge choice.

The structure of this paper is as follows: In Section II, we present the theory of calculating CWFs. Section III 
provides a detailed discussion on the implementation of the method. A series of benchmark calculations are presented 
in Section IV. Finally, in Section V, we summarize the theory of CWFs and suggest a potential role for CWFs as a foundation 
for developing novel methods.

\section{THEORY}

 Let us start by introducing Bloch functions $\{\phi\}$, which can be obtained by solving 
 the Kohn-Sham (KS) equation \cite{Kohn1965} within the density functional theory (DFT) \cite{Hohenberg1964}, normalized as 
 \begin{eqnarray}
    \langle \phi_{{\bf k}_1\mu}\vert\phi_{{\bf k}_2\nu}\rangle
    = N_{\rm BC}\delta_{{\bf k}_1{\bf k}_2}\delta_{\mu\nu},
    \label{eq:Bloch}
 \end{eqnarray}
 where ${\bf k}$ and $\mu$ are the k-vector and the band index, respectively, and 
 $N_{\rm BC}$ is the number of primitive cells in the Born-von Karman (BvK) boundary condition.
 We consider a projection of a localized guiding function $Q$ onto the Bloch functions $\{\phi\}$ weighted 
 with a window function $w(\varepsilon)$ as 
 \begin{eqnarray}
    \vert L_{{\bf R}p} \rangle
     = 
    \frac{1}{N_{\rm BC}}
     \sum_{{{\bf k},\mu}} 
     {\rm e}^{-{\rm i}{\bf k}\cdot{\bf R}}
     \vert \phi_{{\bf k}\mu} \rangle 
     a_{{\bf k}\mu,p}
    \label{eq:proj}
 \end{eqnarray}
 with 
 \begin{eqnarray}
   a_{{\bf k}\mu,p} 
   =
     w(\varepsilon_{{\bf k}\mu})
    \langle\phi_{{\bf k}\mu}\vert Q_{{\bf 0}p}\rangle,
    \label{eq:cdef}
 \end{eqnarray}
 where 
 ${\bf R}$ and $p$ are a translational lattice vector and the index of localized orbital, respectively,
 $\varepsilon_{{\bf k}\mu}$ is the eigenvalue of the KS equation,  
 and $Q_{{\bf 0}p}\equiv Q_{{\bf R}p}~({\bf R}={\bf 0})$.
 In the summation of Eq.~(\ref{eq:proj}), the numbers of k-point and the Bloch functions to be included 
 are $N_{\rm BC}$ and $N_{\rm band}$ per k-point, respectively.
 Throughout the paper we do not consider the spin dependency on the formulation for sake of simplicity, 
 but the generalization is straightforward. 
 The localized functions $\{Q_{{\bf R} p}\}$, such as atomic orbitals, 
 hybrid orbitals, or molecular orbitals (MOs), are assumed to localize in the unit cell specified by ${\bf R}$. 
 When the window function $w(\varepsilon)$ is taken to be unity, Eq.~(\ref{eq:proj}) is nothing but an expansion of 
 $Q_{{\bf R} p}$ with Bloch functions from the identity relation 
 $\frac{1}{N_{\rm BC}}\sum_{{{\bf k}\mu}}\vert \phi_{{\bf k}\mu} \rangle\langle \phi_{{\bf k}\mu} \vert = \hat{I}$.
 To introduce an expansion of $Q_{{\bf R}p}$ by the Bloch functions $\{\phi_{{\bf k}\mu}\}$ in a subspace, 
 we use the following window function $w(\varepsilon)$:
 \begin{eqnarray}
    w(\varepsilon)
   = 
   \frac{1-\exp(x_0+x_1)}{\left(1+\exp(x_0))(1+\exp(x_1)\right)} + \delta
    \label{eq:window}
 \end{eqnarray}
 with the definition of $x_0\equiv\beta(\varepsilon_0-\varepsilon)$ and $x_1\equiv\beta(\varepsilon-\varepsilon_1)$, 
 where $\beta=\frac{1}{k_{\rm B}T}$ with the Boltzmann constant of $k_{\rm B}$ and a temperature of $T$. 
 The window function of Eq.~(\ref{eq:window}) is obtained by subtracting 1 from the sum of two functions of $1/(1+\exp(x))$.
 The last term $\delta$ is a small constant, e.g., $ 10^{-12}$, which is introduced to avoid the ill-conditioning of 
 the matrix consisting of $a_{{\bf k}\mu,p}$ by Eq.~(\ref{eq:cdef}) as discussed later on. 
 As shown in Fig.~1, the parameters of $\varepsilon_0$ and $\varepsilon_1$ ($\varepsilon_0<\varepsilon_1$) 
 determine the range of energy where the Bloch states are included in the expansion of Eq.~(\ref{eq:proj}) 
 with a large weight, and the temperature $T$ gives the degree of smearing around $\varepsilon_0$ 
 and $\varepsilon_1$ in $w(\varepsilon)$. To control the degree of smearing, we introduce the temperature of $T$ 
 so that one can intuitively understand the degree of smearing.  
 It is noted that the choice of the window function by Eq.~(\ref{eq:window}) is not unique, 
 and the other choices should give almost equivalent results as long as they are chosen properly.

\begin{figure}[t]
    \centering
    \includegraphics[width=8.5cm]{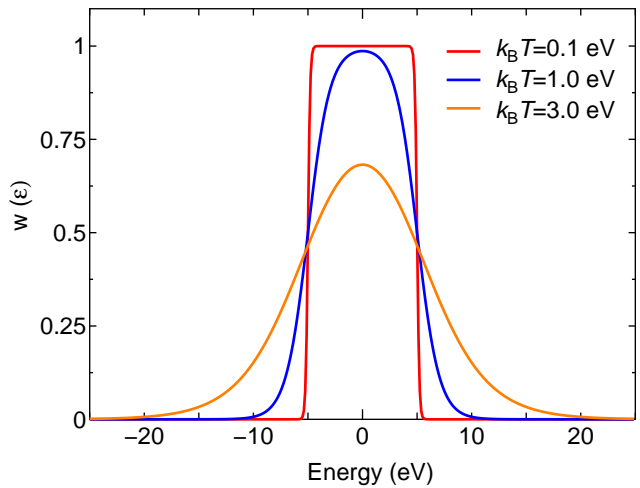}
    \caption{
     Window functions of Eq.~(\ref{eq:window}) with $k_{\rm B}T$=0.1, 1.0, and 3.0 eV. 
     $\varepsilon_0$ and $\varepsilon_1$ were set to be -5.0 and 5.0 eV, respectively, and $\delta=10^{-12}$ was used.
    }
\end{figure}

 The function $L_{{\bf R}p}$ defined by Eq.~(\ref{eq:proj}) must be similar to the localized function $Q_{{\bf R}p}$, 
 but they are non-orthogonal to each other in general. We now consider generating a set of closest Wannier functions (CWFs)
 $\{W_{{\bf R}p}\}$ to a set of the localized functions $\{L_{{\bf R}p}\}$ in a sense that the sum of 
 the squared distance between $L$ and $W$ in the Hilbert space is minimized. 
 As well as Eq.~(\ref{eq:proj}), noting that the CWFs can be expressed \cite{Marzari12} as  
 \begin{eqnarray}
    \vert W_{{\bf R} p} \rangle
     = 
    \frac{1}{N_{\rm BC}}
     \sum_{{{\bf k},\mu}} 
     {\rm e}^{-{\rm i}{\bf k}\cdot{\bf R}}
     \vert \phi_{{\bf k}\mu} \rangle
     b_{{\bf k}\mu,p},
    \label{eq:CWF-expand}
 \end{eqnarray}
 and defining the residual function $R$:
 \begin{eqnarray}
    \vert R_{{\bf R} p} \rangle
     = 
    \vert L_{{\bf R} p}\rangle - \vert W_{{\bf R} p}\rangle,
    \label{eq:residualF}
 \end{eqnarray}
 the distance measure (DM) function $F$ we minimize is given by 
 \begin{eqnarray}
   \nonumber
   F[B]
   &=& 
   \sum_{p}
   \langle R_{{\bf 0} p} \vert R_{{\bf 0} p} \rangle,\\
   &=& 
  \frac{1}{N_{\rm BC}}
  \sum_{{\bf k}}X[B,{\bf k}]
  \label{eq:object}
 \end{eqnarray}
 with 
 \begin{eqnarray}
   X[B,{\bf k}] = 
  {\rm tr}\left[
    \left(A^{\dag}({\bf k})-B^{\dag}({\bf k})\right)
    \left(A({\bf k})-B({\bf k})\right)
          \right],
  \label{eq:tCk}
 \end{eqnarray}
 where the elements of the matrices $A({\bf k})$ and $B({\bf k})$ are given by 
 $a_{{\bf k}\mu,p}$ and $b_{{\bf k}\mu,p}$, respectively. 
 The second line of Eq.~(\ref{eq:object}) is obtained by considering 
 the orthonormality of Eq.~(\ref{eq:Bloch}), and $X[B,{\bf k}]$ of Eq.~(\ref{eq:tCk}) is regarded as 
 the squared Frobenius norm of $\left(A({\bf k})-B({\bf k})\right)$ \cite{Golub1996}. 
 The same Bloch functions as in Eq.~(\ref{eq:proj}) are included in the summation of Eq.~(\ref{eq:CWF-expand}).
 The number of CWFs to be generated is $N_{\rm CWF}$ per unit cell, being equivalent 
 to the number of the guiding functions per cell, and $N_{\rm CWF}$ should be smaller than 
 or equal to $N_{\rm band}$ to guarantee the linear independency of the subspace spanned by the CWFs, 
 resulting in size of $ N_{\rm band}\times N_{\rm CWF}$ for the matrices $A({\bf k})$ and $B({\bf k})$.
Assuming $B^{\dag}({\bf k})B({\bf k})=I$, where $I$ is of size $N_{\rm CWF}\times N_{\rm CWF}$, 
the CWFs are ensured to form a set of orthonormal functions. 
Under this constraint, we consider the optimization of the DM function $F$. 
Furthermore, the matrix $B({\bf k})$ is regarded as a part of a unitary matrix of size $N_{\rm band}\times N_{\rm band}$, 
and will be referred to as a {\it partial} unitary matrix in subsequent discussions.

 The minimum of the DM function of Eq.~(\ref{eq:object}) is obtained
 by choosing $B({\bf k})$ as $U({\bf k})$ which is calculated by the polar decomposition 
 of $A({\bf k})=U({\bf k})P({\bf k})$, 
 where $U({\bf k})$ and $P({\bf k})$ are unitary and hermitian, respectively \cite{Fan1955}. 
 Let us prove the statement below. The polar decomposition of $A({\bf k})$ is obtained via the singular 
 value decomposition (SVD) of $A({\bf k})$ as
 \begin{eqnarray}
   A = W\Sigma V^{\dag} = WV^{\dag}V\Sigma V^{\dag} = UP,
  \label{eq:SVD}
 \end{eqnarray}
 where we dropped the dependency on ${\bf k}$ for simplicity of the notation, and hereafter
 we will denote the dependency if necessary. 
 $W$ and $V$ are the left and right singular matrices in size of 
 $N_{\rm band}\times N_{\rm CWF}$ and $N_{\rm CWF}\times N_{\rm CWF}$, respectively, 
 and $\Sigma$ is the singular value diagonal matrix in size of 
 $N_{\rm CWF}\times N_{\rm CWF}$. 
 Note that $U\equiv WV^{\dag}$ and $P\equiv V\Sigma V^{\dag}$. 
 It is worth mentioning that $W$ and $U$ are partial unitary matrices, and hold $W^{\dag}W=I$ ($WW^{\dag}\neq I$) 
 and $U^{\dag}U=I$ ($UU^{\dag}\neq I$), and that $V$ is a full unitary matrix holding $V^{\dag}V=I$ ($VV^{\dag}=I$).
 We evaluate $X$ of Eq.~(\ref{eq:tCk}) for each ${\bf k}$ with both the matrix $U$ obtained by the polar decomposition 
 and an arbitrary partial unitary matrix $B$, and calculate the difference as 
 \begin{eqnarray}
   \nonumber
   X[U] - X[B]
   &=&
   2{\rm tr}
   \left[\frac{1}{2}\left(A^{\dag}B+B^{\dag}A\right) - P\right],\\
   \nonumber
   &=&
   2{\rm tr}
   \left[\Sigma D - \Sigma\right],\\
   &=& 
   2\sum_{n}\sigma_{n}\left( d_{nn} - 1\right)
  \label{eq:diff_F}
 \end{eqnarray}
 with
 \begin{eqnarray}
   D=\frac{1}{2}\left(V^{\dag}U^{\dag}BV+V^{\dag}B^{\dag}UV\right).
  \label{eq:MatB}
 \end{eqnarray}
 The second line of Eq.~(\ref{eq:diff_F}) is derived by using the polar decomposition of $A$ and $P=V\Sigma V^{\dag}$, 
 and $\sigma_{n}$ and $d_{nn}$ in the third line are diagonal elements of $\Sigma$ and $D$, respectively.
 The upper bound of the diagonal elements of the hermitian matrix $D$ is found to be unity, since the matrix $D$ consists of 
 the sum of the product of partial unitary matrices of $V^{\dag}U^{\dag}$ ($V^{\dag}B^{\dag}$) and $BV$ ($UV$). 
 Another proof for the upper bound is also given based on the Cauchy-Schwarz inequality in the appendix. 
 By considering the upper bound of the diagonal elements of $D$ and $0 \leq\sigma_n$, 
 the third line of Eq.~(\ref{eq:diff_F}) leads to the following inequality:
 \begin{eqnarray}
    X[U] \leq X[B].
    \label{eq:inequality}
 \end{eqnarray}
 The equality of Eq.~(\ref{eq:inequality}) holds if $B=U$. 
 If some of $\sigma_n$ is zero, the corresponding $d_{nn}$ in Eq.~(\ref{eq:diff_F}) can be arbitrarily chosen. 
 So, we see from Eq.~(\ref{eq:MatB}) that $U$ is not uniquely determined. 
 If all the singular values of $\sigma_n$ are positive, all the $d_{nn}$s should be unity when $X[U]=X[B]$.
 The uniqueness of the solution can be confirmed as follows:
 Since the matrix $D$ consists of the products of the partial unitary matrices as discussed above, 
 the case that all the diagonal elements $d_{nn}$ are unity happens when 
 \begin{eqnarray}
    U^{\dag}B + B^{\dag}U = 2 I,
    \label{eq:UC}
 \end{eqnarray}
 which is derived by equating $D=I$ and multiplying $V$ and $V^{\dag}$ from the left and right of the equation, respectively.  
 By noting again that $B$ and $U$ in Eq.~(\ref{eq:UC}) are the partial unitary matrices, it is found that Eq.~(\ref{eq:UC}) 
 holds if and only if $B=U$.
 Thus, we have proven the statement that the minimum of the DM function of Eq.~(\ref{eq:object}) is 
 uniquely determined when $B=U$ as long as all the singular values of $\sigma_n$ are positive, 
 since $F[B]$ of Eq.~(\ref{eq:object}) consists of the sum of $X[B,{\bf k}]$ over {\bf k}.
 The uniqueness of $U$ itself is related to that of the SVD for the matrix $A$ of Eq.~(\ref{eq:cdef}). 
 If $A$ has $N_{\rm CWF}$ positive singular values which are non-degenerate, the SVD is uniquely determined 
 except for the non-unique phases in $W$ and $V$. 
 Even if there are degenerate singular values, the matrix $U$ is invariant, since the freedom of the unitary 
 transformation $K$ for the degenerate left and right singular vectors is canceled out as 
 $U=WKK^{\dag}V^{\dag}=WV^{\dag}$ with $KK^{\dag}=I$.
 Further, noting that the non-unique phases in $W$ and $V$ are canceled out when 
 the matrix $U$ is computed as $WV^{\dag}$, we conclude that $U$ is uniquely determined if $A$ has the $N_{\rm CWF}$ positive 
 singular values. The violation from the positive definiteness of the singular values can be avoided 
 by the small constant of $\delta$ in Eq.~(\ref{eq:window}). 
 On the other hand, if $\delta$ in Eq.~(\ref{eq:window}) is set to be 0 and a narrow window,
 including fewer than $N_{\rm CWF}$ eigenstates for a given ${\bf k}$, is used with a small $k_{\rm B}T$, 
 the matrix $A$ has $N_{\rm PSV}$ positive singular values, where $N_{\rm PSV}<N_{\rm CWF}$. 
 In this case, a subspace of the $\left(N_{\rm CWF}-N_{\rm PSV}\right)$ dimension 
 is arbitrary chosen to form $W$ and $V$ of the dimension of $N_{\rm CWF}$ in addition to the subspace of the $N_{\rm PSV}$ dimension, 
 which breaks the symmetry of CWFs.  
 The situation can be avoided by introducing a small constant $\delta$ in the window function of Eq.~(\ref{eq:window}). 
 The simple treatment restores the symmetry of CWFs, allowing us to calculate symmetry preserving CWFs 
 for a wide variety of choices in the energy range of the window function.
 It is also noted that the non-unique phase of the Bloch functions is canceled out via the polar decomposition of $A$ 
 and Eq.~(\ref{eq:CWF-expand}). 
 Thus, we see that the proposed method is free from complications arising from the choice of gauge. 

 Using Eqs.~(\ref{eq:tCk}) and (\ref{eq:SVD}), the minimum of the DM function $F$ at $B=U$ is given by 
 \begin{eqnarray}
   F[U]
   =
  \frac{1}{N_{\rm BC}}
  \sum_{{\bf k},p}
   \left(\sigma_{{\bf k} p}-1\right)^2,
   \label{eq:object2}
 \end{eqnarray}
 where $\sigma_{{\bf k} p}$ is a singular value of the matrix $A({\bf k})$.
 From Eq.~(\ref{eq:object2}), we find that the mean squared distance between $L$ and $W$ is related to
 the deviation of $\sigma$ from unity. 

 The proposed method based on the polar decomposition of $A$ is closely related to the L\"{o}wdin orthogonalization \cite{Lowdin1950}
 as shown below.
 The Fourier transform of the overlap integrals for $\{L\}$ is given by 
 \begin{eqnarray}
  \sum_{{\bf R}}
  {\rm e}^{{\rm i}{\bf k}\cdot{\bf R}}
   \langle L_{{\bf 0}p}\vert L_{{\bf R}q}\rangle
   =
   \sum_{\mu}a_{{\bf k}\mu,p}^{*}a_{{\bf k}\mu,q},
   \label{eq:FT_OL_L}
 \end{eqnarray}
 which is obtained by noting that 
 $\frac{1}{N_{\rm BC}}\sum_{{\bf R}}{\rm e}^{{\rm i}({\bf k}-{\bf k}')\cdot{\bf R}}=\delta_{{\bf k}{\bf k}'}$.
 By writing Eq.~(\ref{eq:FT_OL_L}) as $S({\bf k}) = A^{\dag}({\bf k})A({\bf k})$ in a matrix form, 
 and using Eq.~(\ref{eq:SVD}), we have the following relation:
 \begin{eqnarray}
  S({\bf k}) = A^{\dag}({\bf k})A({\bf k})
  = 
  V^{\dag}({\bf k})\Sigma^2({\bf k}) V({\bf k}).
   \label{eq:SL1}
 \end{eqnarray}
 Comparing Eq.~(\ref{eq:SL1}) with $P({\bf k})=V({\bf k})\Sigma({\bf k})V^{\dag}({\bf k})$, 
 one obtains a relation $P({\bf k})=S^{1/2}({\bf k})$. If $P({\bf k})$  is invertible, the matrix $U({\bf k})$
 is given with Eq.~(\ref{eq:SVD}) by 
 \begin{eqnarray}
  U({\bf k}) = A({\bf k})S^{-1/2}({\bf k}).
   \label{eq:U_IS}
 \end{eqnarray}
 The matrix $U$ obtained by Eq.~(\ref{eq:U_IS}) is exactly equivalent to that by the L\"{o}wdin orthogonalization \cite{Lowdin1950}, and  
 the closest property of the L\"{o}wdin orthogonalized orbitals to a given set of orbitals was proven in Ref.~\cite{Carlson57}. 
 On the other hand, the proposed method does not require the calculation of $S^{-1/2}$, and can be applied
 in a numerical stable manner even to the case that the matrix $A$ is nearly ill-conditioned. 
 The definition of $A$ by Eq.~(\ref{eq:cdef}) with the window function allows us to calculate CWFs 
 in various respects, e.g., heavily weighting to occupied states and disentangling of localized states 
 embedded in the wider bands as demonstrated in Section IV. 
 In this sense, the method we propose can be regarded as a generalization of the L\"{o}wdin orthogonalization.
 
 The disentanglement of bands can be easily performed by properly selecting the window function of Eq.~(\ref{eq:cdef}). 
 Since the $N_{\rm band} (\ge N_{\rm CWF})$ Bloch functions per k-point are included in Eqs.~(\ref{eq:proj}) and (\ref{eq:CWF-expand}),
 it should be emphasized that the proposed method always disentangles $N_{\rm band}$ states to generate $N_{\rm CWF}$ Wannier functions. 
 A couple of examples will be shown in Section IV, including valence and low-lying conduction bands for
 the diamond Si and narrow $3d$-bands embedded in the wider $4s$-band for the face centered cubic (FCC) Cu. 

 Once the $k$-dependent $U({\bf k})$ are obtained by the polar decomposition, tight-binding (TB) parameters are
 calculated as expectation values of the KS Hamiltonian $\hat{H}_{\rm KS}$ by summing contributions over ${\bf k}$ and $\mu$ as 
 \begin{eqnarray}
   \nonumber
   t_{{\bf 0}p,{\bf R}q}
    &=&
   \langle W_{{\bf 0}p}\vert \hat{H}_{\rm KS} \vert W_{{\bf R}q}\rangle,\\
    &=&
    \sum_{{\bf k},\mu} \varepsilon_{{\bf k}\mu} u_{{\bf k}\mu,p}^{*} u_{{\bf k}\mu,q} {\rm e}^{-{\rm i}{\bf k}\cdot{\bf R}},
    \label{eq:TB_Para}
 \end{eqnarray}
 where $u_{{\bf k}\mu,q}$ is the matrix element of $U({\bf k})$. 
 Like with MLWFs, the TB parameters can be used for the Wannier interpolation in calculations of band structures and physical quantities. 

The computational procedure to generate the CWFs is summarized as follows:
 \begin{enumerate}

 \item Determining $\varepsilon_0$, $\varepsilon_1$, and $k_{\rm B}T$ in Eq.~(\ref{eq:window}) 
 by checking the band dispersion of a system of interest. 
 The parameters should be chosen properly so that the targeted eigenstates can be included. 

 \item Choosing a set of localized orbitals \{$Q$\}.
 Atomic orbitals, hybrid orbitals, and MOs might be possible choices.
 Depending on the symmetry of the targeted eigenstates, a proper set of localized orbitals 
 needs to be chosen, e.g., $d$-orbitals need to be employed in the generation of CWFs 
 for the $d$-bands in FCC Cu as shown later on. 

 \item Calculation of $A({\bf k})$ by Eq.~(\ref{eq:cdef}). 
  The calculation of the overlap $\langle\phi_{{\bf k}\mu}\vert Q_{{\bf 0} p}\rangle$ 
  depends on the implementation of the KS method. 
  
 \item Performing the SVD of $A({\bf k})$.
  A proper numerical library might be used. 

 \item Calculation of $U({\bf k})$.
 Using the left and right singular matrices, $W({\bf k})$ and $V({\bf k})$, of $A({\bf k})$,  
 $U({\bf k})$ is calculated as $W({\bf k})V^{\dag}({\bf k})$. 

 \item Calculations of the CWFs and TB parameters.
 The CWFs can be calculated using Eq.~(\ref{eq:CWF-expand}) with $B=U$ obtained 
 by the step 5. The calculation is expressed by the sum of the matrix-matrix product over ${\bf k}$.
 Also, the TB parameters are obtained by Eq.~(\ref{eq:TB_Para}).

 \end{enumerate}
 It should be stressed that the minimization of $F$ can be efficiently performed in a non-iterative manner
 by the steps 1 to 6 using the polar decomposition via the SVD. 
 The computational cost of each step is estimated to be 
 $O(N_{\rm BC}N_{\rm band}N_{\rm CWF}N_{\rm basis})$,
 $O(N_{\rm BC}N_{\rm band}^3)$, 
 $O(N_{\rm BC}N_{\rm band}N_{\rm CWF}^2)$, 
 or 
 $O(N_{\rm BC}N_{\rm band}N_{\rm CWF}N_{\rm basis})$
 for the step 3, 4, 5, or 6, respectively, where $N_{\rm basis}$ is the number of basis functions to expand the KS orbitals. 
 If the localized basis functions are used, the computational cost of the step 3 becomes  
 $O(N_{\rm BC}N_{\rm band}N_{\rm CWF})$. 

 For isolated systems, the Brillouin zone sampling is limited to only the $\Gamma$ point. 
 No modification of the proposed method is needed to calculate the CWFs.

 The theoretical framework we have discussed so far is general for any implementation of the DFT-KS method. 
 However, the choice of the the localized functions $\{Q\}$ in Eq.~(\ref{eq:cdef}) may depend on the implementation.
 So, we will discuss how the localized functions $\{Q\}$ can be properly generated in Section III.

\section{IMPLEMENTATION}

\subsection{General}

We have implemented the proposed method into the OpenMX DFT software package \cite{OpenMX,Ozaki2005,Duy2014} which is based on norm-conserving 
pseudopotentials (PPs) \cite{MBK1993,Theurich2001} and optimized pseudo-atomic orbitals (PAOs) \cite{Ozaki2003,Ozaki2004} as basis set. 
All the benchmark calculations were performed with a computational condition of a production level. 
The basis functions used are listed in Table~\ref{table:basis}.
In the abbreviation of basis functions such as H7.0-s2p2d1, H stands for the atomic symbol, 
7.0 the cutoff radius (Bohr) in the generation by the confinement scheme \cite{Ozaki2003,Ozaki2004}, 
and s2p2d1 means the employment of two, two, 
and one optimized radial functions for the $s$-, $p$-, and $d$-orbitals, respectively.
The radial functions were optimized by a variational optimization method \cite{Ozaki2003}. 
These basis functions we used can be regarded as at least double zeta plus double polarization basis sets 
if we follow the terminology of Gaussian or Slater-type basis functions. 
Valence states included in the PPs are listed in Table~\ref{table:basis}.
All the PPs and PAOs we used in the study were taken from the database (2019) in the OpenMX website \cite{OpenMX}, which 
were benchmarked by the delta gauge method \cite{Lejaeghere2016}. 
Real space grid techniques were used for the numerical integrations 
and the solution of the Poisson equation using fast Fourier transform (FFT) 
with the energy cutoff of 250 to 500 Ryd \cite{Ozaki2005}.
We used a generalized gradient approximation (GGA) proposed by Perdew, Burke, and Ernzerhof
to the exchange-correlation functional \cite{Perdew1996}. An electronic temperature of 300 K was used to count the number
of electrons by the Fermi-Dirac function for all the systems we considered.
For all the calculations, $\delta=10^{-12}$ was used in the window function of Eq.~(\ref{eq:window}), and 
$N_{\rm band}$ was set to be equilavent to the number of basis functions in both the summation of Eqs.~(\ref{eq:proj}) 
and (\ref{eq:CWF-expand}). 

   \begin{table}[t]
    \caption{
     Basis functions and valence states included in PPs. 
     $^*$For the calculations of atomic charges, the basis functions listed in Table~\ref{table:Charges} were used. 
    }
   \vspace{3mm}
   \begin{tabular}{ccccccc}
   \hline\hline
    Element     &&& Basis functions &&&  Valence states in PP\\
     \hline
      H    &&&   H7.0-s2p2d1    &&& $1s$  \\
      C    &&&   C6.0-s3p2d2    &&& $2s$, $2p$  \\
      N    &&&   N6.0-s3p2d2    &&& $2s$, $2p$  \\
      Na   &&&  $^*$Na9.0-s3p3d2f2 &&& $2s$, $2p$, $3s$   \\
      Si   &&&   Si7.0-s2p2d2f1 &&& $3s$, $3p$  \\
      Cl   &&&  $^*$Cl9.0-s3p3d2f2 &&& $3s$, $3p$  \\
      S    &&&   S7.0-s3p2d2f1  &&& $3s$, $3p$  \\
      Cu   &&&   Cu6.0-s3p3d3f1 &&& $3s$, $3p$, $3d$, $4s$ \\
      Se   &&&   Se7.0-s3p2d2   &&& $4s$, $4p$  \\
      Bi   &&&   Bi8.0-s3p2d2f1 &&& $5d$, $6s$, $6p$ \\
   \hline
   \end{tabular}
   \label{table:basis}
  \end{table}

\subsection{Choice of Guiding Functions $\{Q\}$}

Depending on the targeted eigenstates in the window function of Eq.~(\ref{eq:window}), 
one can choose either atomic orbitals, hybrid atomic orbitals, or MOs as the guiding functions $\{Q\}$ in Eq.~(\ref{eq:cdef}). 
In this subsection we discuss the three kinds of the localized guiding functions $\{Q\}$, i.e., atomic orbitals, hybrid atomic orbitals, 
and embedded MOs in molecules and bulks, and how they can be generated in our implementation. 

In our implementation, the Bloch function $\phi_{{\bf k}\mu}$ is expanded by PAOs $\chi$ \cite{Ozaki2003,Ozaki2004} as 
 \begin{eqnarray}
    \vert \phi_{{\bf k}\mu}\rangle
     = 
    \sum_{{\bf R}}
    {\rm e}^{{\rm i}{\bf k}\cdot {\bf R}}
    \sum_{i\alpha} 
    c_{ {\bf k} \mu,i\alpha} 
    \vert \chi_{{\bf R} i\alpha} \rangle,
   \label{eq:LCPAO}
 \end{eqnarray}
 where $i$ and $\alpha$ are atomic and orbital indices, and $c$ is a linear combination of PAO (LCPAO) coefficient. 
 Also note that $\langle {\bf r}\vert \chi_{{\bf R} i\alpha} \rangle \equiv \chi_{i\alpha}({\bf r}-{\bf \tau}_{i}-{\bf R})$, 
 where ${\bf \tau}_{i}$ is the position of atom $i$. 
 We use PAOs as the guiding functions $\{Q\}$ of atomic orbitals, corresponding to valence orbitals or 
 specific orbitals such as localized $d$-orbitals in a narrow energy window. 

 The choice of the atomic orbitals gives a good guiding function, however, 
 the resultant CWFs may not recover the symmetry of CWFs respecting the bond direction 
 to the neighboring atoms unlike hybrid atomic orbitals due to the closest property of CWFs.
 In this case, it would be better to use the hybrid atomic orbitals rather than the atomic orbitals as explained below. 
 Let us introduce a projection operator for the occupied space by
 \begin{eqnarray}
    \hat{P}
    =
    \frac{1}{N_{\rm BC}}
    \sum_{{\bf k} \mu}
    \vert \phi_{\bf k \mu}\rangle
     f(\varepsilon_{\bf k \mu})
    \langle \phi_{\bf k \mu}\vert,
    \label{eq:PO}
 \end{eqnarray}
 where $f$ is the Fermi-Dirac function. With the projection operator, the density matrix is calculated by 
 \begin{eqnarray}
   \nonumber
   \rho_{{\bf R} i\alpha,{\bf R}' j\beta}
    &=& 
    \sum_{{\bf k} \mu}
    \langle \widetilde{\chi}_{{\bf R} i\alpha} \vert \hat{P}
    \vert \widetilde{\chi}_{{\bf R}' j\beta} \rangle,\\
   \nonumber
    &=&
    \frac{1}{N_{\rm BC}}
    \sum_{{\bf k} \mu}
    {\rm e}^{{\rm i}{\bf k}\cdot \left({\bf R}-{\bf R}'\right)}
    f(\varepsilon_{\bf k \mu})
    c_{ {\bf k} \mu,i\alpha} 
    c^{*}_{ {\bf k} \mu,j\beta},\\
    \label{eq:DM}
 \end{eqnarray}
 where $\widetilde{\chi}$ is the dual orbital defined by 
 \begin{eqnarray}
    \vert \widetilde{\chi}_{{\bf R} i\alpha} \rangle
    &=&
    \sum_{{\bf R}' j\beta}
    \vert \chi_{{\bf R}' j\beta} \rangle
     S_{{\bf R}'j\beta, {\bf R}i\alpha}^{-1}
    \label{eq:dualorb}
 \end{eqnarray}
 holding the following orthonormality:
 \begin{eqnarray}
  \langle \chi_{{\bf R} i\alpha} \vert \widetilde{\chi}_{{\bf R}' j\beta} \rangle 
   &=& \langle \widetilde{\chi}_{{\bf R} i\alpha} \vert \chi_{{\bf R}' j\beta} \rangle = \delta_{{\bf RR}'}\delta_{ij}\delta_{\alpha
    \label{eq:ortho_dual}
\beta}.
 \end{eqnarray}
 By setting ${\bf R}={\bf R}'={\bf 0}$ and $i=j$ in Eq.~(\ref{eq:DM}), and diagonalizing the diagonal block element 
 consisting of $\rho_{{\bf 0} i\alpha,{\bf 0} i\beta}$, 
 which is associated with selected atomic orbitals on the atomic site $i$, we obtain the hybrid 
 atomic orbitals respecting the symmetry of the bond direction to the neighboring atoms. 
 We employ the hybrid atomic orbitals as the localized functions $\{Q\}$ in our implementation.
 When the same atomic orbitals are chosen to form the diagonal block element as for the case 
 of atomic orbitals $\{Q\}$, the resultant CWFs span the same subspace in the Hilbert space. 

 For some systems the electronic structures can be rather easily understood by employing MOs as building blocks. 
 A molecular crystal is such a case, where the eigenstates near the chemical potential can be approximately constructed by a linear
 combination of the MOs of constituting molecules. Another example is the bond in molecules and bulks. 
 The bond between two atoms embedded in the system might be analyzed by MOs associated with the two atoms. 
 Here we show how embedded MOs in molecules and bulks can be calculated in a simple procedure.
 Let us start by noting that using Eq.~(\ref{eq:PO}) and assuming that the Bloch function is expressed by Eq.~(\ref{eq:LCPAO})
 the total number of electrons for a spin degenerate case is given by \cite{Ozaki2018}
 \begin{eqnarray}
  \nonumber
  N_{\rm ele}
  &=&
  2{\rm tr}
  \left[
   \hat{P}
  \right],\\
  &=&
  2\sum_{{\bf R} i\alpha}
  \langle \widetilde{\chi}_{{\bf R} i\alpha} \vert 
  \hat{P}
  \vert \chi_{{\bf R} i\alpha} \rangle,
  \label{eq:P_Nele}
 \end{eqnarray}
 In the second line of Eq.~(\ref{eq:P_Nele}) we used the following identity operator:
 \begin{eqnarray}
   \hat{I}
    =
    \sum_{{\bf R} i\alpha}
    \vert \chi_{{\bf R} i\alpha} \rangle
    \langle \widetilde{\chi}_{{\bf R} i\alpha} \vert 
    =
    \sum_{{\bf R} i\alpha}
    \vert \widetilde{\chi}_{{\bf R} i\alpha} \rangle
    \langle \chi_{{\bf R} i\alpha} \vert. 
    \label{eq:iden}
 \end{eqnarray}
 Since $N_{\rm ele}=N_{\rm BC}N_{\rm ele}^{(0)}$ with  
 $N_{\rm ele}^{(0)}=2\sum_{i\alpha}\langle \widetilde{\chi}_{{\bf 0} i\alpha} \vert \hat{P} \vert \chi_{{\bf 0} i\alpha} \rangle$, 
 it is enough to consider $N_{\rm ele}^{(0)}$. We utilize the formula for $N_{\rm ele}^{(0)}$
 to calculate embedded MOs in molecules and bulks, and rewrite it the sum of partial traces as 
 \begin{eqnarray}
   N_{\rm ele}^{(0)}
    =
    2\sum_{g}
    {\rm tr}_{g}
    \left[
    (\widetilde{\chi}_{{\bf 0}g}\vert 
     \hat{P} 
     \vert \chi_{{\bf 0}g})\right]
    = 
    2\sum_{g}
    {\rm tr}_{g}
    \left[
    \Lambda_{g}
    \right] 
    \label{eq:SumTr},
 \end{eqnarray}
 where $g$ is the index of group including PAOs on grouped atoms, e.g., a set of PAOs on a molecule.
 The notation of $\vert \chi_{{\bf 0}g})$ stands for a subset of PAOs as 
 \begin{eqnarray}
   \vert \chi_{{\bf 0}g} )
   = 
   \left(
    \cdots,
     \vert \chi_{{\bf 0}i1}\rangle,
     \vert \chi_{{\bf 0}i2}\rangle,
     \cdots,
   \right),
    \label{eq:setchi}
 \end{eqnarray}
 where PAOs on all the atoms in the group $g$ are included in the subset. 
 The notation of $\vert \widetilde{\chi}_{{\bf 0}g})$ is the counterpart for the dual orbitals.
 Using the identity operator of Eq.~(\ref{eq:iden}), $\Lambda_{g}$ in Eq.~(\ref{eq:SumTr}) is given by 
 \begin{eqnarray}
   \Lambda_{g}
   = 
   \sum_{{\bf R}g'}
   \rho_{{\bf 0}g,{\bf R}g'} S_{{\bf R}g',{\bf 0}g}
  \label{eq:Lambda}
 \end{eqnarray}
 with definition of block elements:
 \begin{eqnarray}
   \rho_{{\bf R}g,{\bf R}'g'}
   &=& 
   (\widetilde{\chi}_{{\bf R}g}\vert \hat{P} \vert \widetilde{\chi}_{{\bf R}'g'} ),
   \label{eq:blockRho}\\
   S_{{\bf R}g,{\bf R}'g'}
   &=& 
   (\chi_{{\bf R}g}\vert \chi_{{\bf R}'g'} ).
    \label{eq:blockS}
 \end{eqnarray}
 We now introduce the {\it embedded} MOs orbitals in molecules and bulks by a right-singular vectors of $\Lambda_{g}$.
 Since the elements of $\Lambda_{g}$ are real in case of the collinear DFT, the right-singular vectors can be 
 obtained with real components. However, the SVD of $\Lambda_{g}$ tends to produce the right-singular vectors with
 complex components, which results in the complex CWFs making analysis complicated. 
 To obtain the real right-singular vectors, we perform the eigendecomposition of $\Lambda_{g}^{\dag}\Lambda_{g}$ as 
 \begin{eqnarray}
   \Lambda_{g}^{\dag}\Lambda_{g}
   = Y_{g}\Omega_{g}^2 Y_{g}^{\dag}, 
   \label{eq:ED_Lambda2}
 \end{eqnarray}
 where $\Omega_{g}^2$ is the diagonal matrix consisting of the eigenvalues of $\Lambda_{g}^{\dag}\Lambda_{g}$, and 
 $Y_{g}$ is a unitary matrix consisting of the corresponding eigenvectors $\{y_{g,\nu}\}$.
 By noting that the SVD of $\Lambda_{g}$ is given by $Z_{g}\Omega_{g} Y_{g}^{\dag}$, 
 we see that the right-singular vectors $Y_{g}$ can be obtained by diagonalizing $\Lambda_{g}^{\dag}\Lambda_{g}$ as $Y_{g}$.
 The square roots of the eigenvalues of $\Lambda_{g}^{\dag}\Lambda_{g}$ are singular values of $\Lambda_{g}$, and 
 can be approximately regarded as the occupation of the corresponding eigenvector $y_{g,\nu}$.
 If $\Lambda_{g}$ is a real matrix, $Y_{g}$ obtained by the diagonalization of $\Lambda_{g}^{\dag}\Lambda_{g}$
 is guaranteed to be a real unitary matrix. We further normalize the eigenvector $y_{g,\nu}$ by considering 
 the overlap integrals of Eq.~(\ref{eq:blockS}) as 
 \begin{eqnarray}
   \vert \bar{y}_{g,\nu}\rangle = \frac{\vert y_{g,\nu}\rangle}
                                   {\sqrt{\langle y_{g,\nu} \vert S_{{\bf 0}g,{\bf 0}g} \vert y_{g,\nu}\rangle}},
   \label{eq:normalized_y}
 \end{eqnarray}
 and calculate an expectation value of the KS Hamiltonian $\hat{H}_{\rm KS}$ with $\bar{y}_{g,\nu}$ as 
 \begin{eqnarray}
   \epsilon_{g,\nu} =  
   \langle \bar{y}_{g,\nu} \vert \hat{H}_{\rm KS} \vert \bar{y}_{g,\nu}\rangle.
   \label{eq:eignormalized_y}
 \end{eqnarray}
 After $\{\bar{y}_{g,\nu}\}$ are ordered based on the expectation values, we employ selected ones among $\{\bar{y}_{g,\nu}\}$, 
 e.g., ones near the chemical potential, as the guiding functions $\{Q\}$ in Eq.~(\ref{eq:cdef}) 
 by monitoring the expectation values and the corresponding singular values.
 They are what we call {\it embedded} MOs in molecules and bulks.
 As demonstrated in Section IV, $\{\bar{y}\}$ work as the good guiding functions to capture  
 the electronic structure of a molecular crystal.

\section{NUMERICAL RESULTS}

\subsection{Wannier Interpolated Bands}

\begin{figure}[t]
    \centering
    \includegraphics[width=8.5cm]{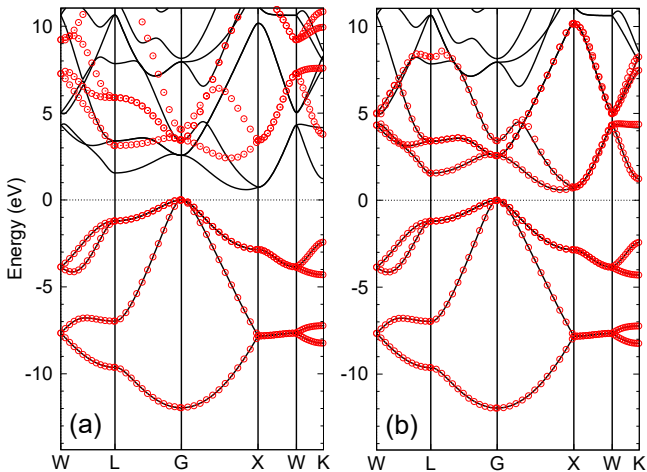}
    \caption{
     Interpolated bands (red circles) of silicon in the diamond structure, calculated by the TB Hamiltonian 
     derived from CWFs with 
     (a) $k_{\rm B}T$=0.01 and (b) 3.0 eV in the window function of Eq.~(\ref{eq:window}), respectively. 
     $\varepsilon_0$ and $\varepsilon_1$ relative to the chemical potential were set to be -15.0 and 0.0 eV, 
     respectively.
     The solid line is the original one directly calculated by the conventional scheme.
     The number of k-points for the Brillouin zone sampling was 13 $\times$ 13 $\times$ 13.
     The experimental lattice constant of 5.43~\AA~was used. 
     The values of the DM function per CWF are 0.571 and 0.444 for (a) and (b), respectively.
    }
   \label{fig:Si-band}
\end{figure}

\begin{figure}[t]
    \centering
    \includegraphics[width=8.0cm]{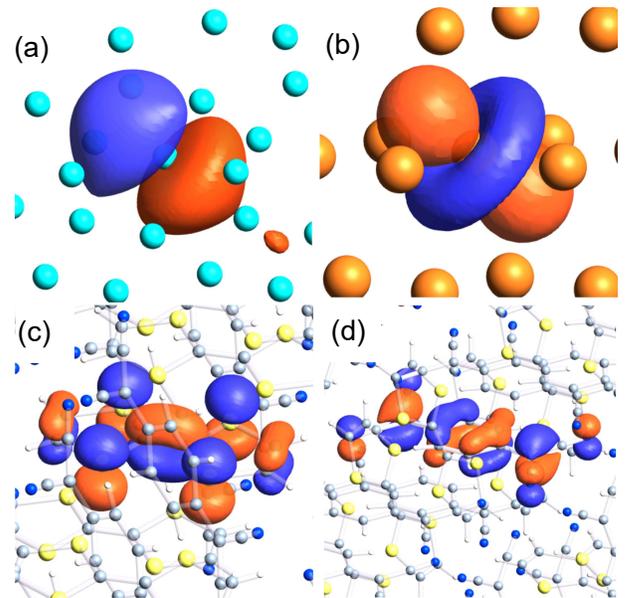}
    \caption{
     CWFs for (a) Si, (b) Cu, (c) TTF in TTF-TCNQ, and (d) TCNQ in TTF-TCNQ.
     In all the cases, isovalues of $\pm 0.04$ (orange:0.04, blue:-0.04) are used for drawing the isosurfaces 
     using OpenMX Viewer \cite{Lee2019}. 
     The computational conditions for (a), (b), (c), and (d) are the same as those in Fig.~\ref{fig:Si-band} (b),
     Fig.~\ref{fig:Cu-band} (a), Fig.~\ref{fig:TTF-TCNQ-band} (b), and Fig.~\ref{fig:TTF-TCNQ-band} (b), respectively.    
    }
   \label{fig:CWFs}
\end{figure}

We present four numerical examples to demonstrate the broad applicability of the proposed method across various systems.

In Figs.~\ref{fig:Si-band} (a) and (b), the interpolated bands of Si in the diamond structure, calculated by the TB Hamiltonian
derived from CWFs, are compared with those by the conventional calculation. 
The hybrid atomic orbitals consisting of valence minimal orbitals, which are one $s$- and a set of $p_x$-, $p_y$-, and $p_z$-orbitals per Si,
were used as the guiding functions $\{Q\}$.
For the window function of Eq.~(\ref{eq:window}) we use -15.0 and 0.0, relative to the chemical potential, 
in eV for $\varepsilon_0$ and $\varepsilon_1$, respectively, which covers the energy range of valence bands. 
On the other hand, two $k_{\rm B}T$ of 0.01 and 3.0 eV were used to check how conduction bands are reproduced depending on 
the effect of smearing. As clearly seen, the two cases well reproduce the valence bands, and the conduction bands 
are also reproduced well in the case of $k_{\rm B}T=3.0$ eV. 
Evaluating the DM function per CWF using Eq.~(\ref{eq:object2}), we obtain values of 0.571 and 0.444 for $k_{\rm B}T$ 
of 0.01 and 3.0 eV, respectively. For $k_{\rm B}T=0.01$ eV, the first four singular values at each $k$ point are close to 1, 
while the remaining four approach $\delta (=10^{-12})$. Conversely, for $k_{\rm B}T=3.0$ eV, the eight singular values 
at each $k$ point distribute ranging from 1.2 to 0.1. 
These differences in the distribution of singular values explain why the value of the DM function for $k_{\rm B}T=0.01$ eV 
is larger than that for $k_{\rm B}T=3.0$ eV.
One of the obtained CWFs is shown in Fig.~\ref{fig:CWFs} (a), which represents a $p$-like CWF.  
In case of $k_{\rm B}T=0.01$ eV, the conduction bands are treated less importantly due to heavily weighting to the valence bands, 
resulting in the poor reproduction of the conduction bands. 
At first glance, one may consider that this is an undesirable feature in such a treatment. 
However, the feature enables us to generate the minimal atomic-like orthogonal orbitals, which well span the subspace 
by the valence bands, and to calculate an effective charge allocated to each atom using the minimal atomic-like orthogonal 
orbitals in an unbiased way. We will demonstrate the calculation of the effective atomic charges and stress the usefulness of the method later on.

\begin{figure}[t]
    \centering
    \includegraphics[width=8.5cm]{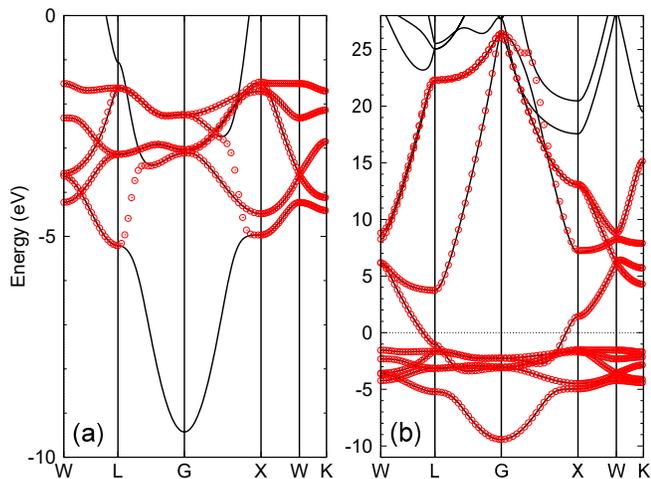}
    \caption{
     Interpolated bands (red circles) of copper in the FCC structure, calculated by the TB Hamiltonian derived 
     from CWFs with
     (a) $\varepsilon_0=-5.5$, $\varepsilon_1=-1.0$ eV, $k_{\rm B}T=1.0$ eV, and hybrid $3d$-orbitals as $\{Q\}$, 
     and 
     (b) $\varepsilon_0=-11.0$, $\varepsilon_1=24.0$ eV, $k_{\rm B}T=3.0$ eV, and hybrid $3d$-, $4s$-, $4p$-orbitals as $\{Q\}$. 
     The solid line is the original one directly calculated by the conventional scheme. 
     The number of k-points for the Brillouin zone sampling was 21 $\times$ 21 $\times$ 21.
     The experimental lattice constant of 3.61~\AA~was used. 
     The values of the DM function per CWF are 0.081 and 0.211 for (a) and (b), respectively.
    }
   \label{fig:Cu-band}
\end{figure}

The disentanglement of bands in metals is demonstrated for Cu in the FCC structure as shown 
in Figs.~\ref{fig:Cu-band} (a) and (b).  
By selecting the parameters $\varepsilon_0$ and $\varepsilon_1$, so that the $3d$ bands are included, 
and only five $d$-orbitals as the guiding functions $\{Q\}$, the five $d$-bands are reproduced 
as shown in Fig.~\ref{fig:Cu-band} (a). We see that the $3d$-bands are properly disentangled
from the dispersive $4s$ band, and one of the obtained CWFs preserves the shape of $d_{z^2}$-orbital 
as shown in Fig.~\ref{fig:CWFs} (b). 
When $3d$-, $4s$-, and $4p$-orbitals are used as the guiding functions $\{Q\}$, a broad range of bands
are reproduced as shown in Fig.~\ref{fig:Cu-band} (b). 
Since our PP of Cu includes $3s$, $3p$, $3d$, and $4s$ states, the original band structure has the deep 
$3s$- and $3p$-bands. We set the parameters in the window function so as to discard the $3s$- and $3p$-bands, and 
use the $3d$-, $4s$-, and $4p$-orbitals, not $3s$- and $3p$-orbitals, as the guiding functions $\{Q\}$. 
So, the $3s$- and $3p$-bands are not included in the interpolated bands (not shown). 
The examples show that the CWFs can be flexibly and easily constructed in accordance with the purpose of study 
without numerical difficulties. 

   \begin{table}[b]
    \caption{
     Expectation values $\epsilon$ (eV) of the KS Hamiltonian calculated by Eq.~(\ref{eq:eignormalized_y}),
     which is relative to the chemical potential, and singular values $\omega$ of $\Lambda_{g}$
     for embedded MOs in the TTF-TCNQ molecular crystal.
    }
   \vspace{3mm}
   \begin{tabular}{ccccccccc}
   \hline\hline
    & TTF &&&&& TCNQ\\
   \hline
     index  & $\epsilon$ & $\omega$ &&& index & $\epsilon$ & $\omega$\\
     \hline
      23   & -0.069  & 1.359 &&& 34 & 0.135 & 1.082\\
      24   &  0.719  & 1.346 &&& 35 & 0.375 & 1.329\\
      25   &  0.761  & 1.449 &&& 36 & 0.477 & 1.336\\
      26   &  3.408  & 0.864 &&& 37 & 4.928 & 0.535\\
      27   &  5.947  & 0.047 &&& 38 & 9.389 & 0.036\\
      28   &  6.197  & 0.047 &&& 39 & 9.795 & 0.019\\
   \hline
   \end{tabular}
   \label{table:MO_data}
  \end{table}

\begin{figure}[t]
    \centering
    \includegraphics[width=8.5cm]{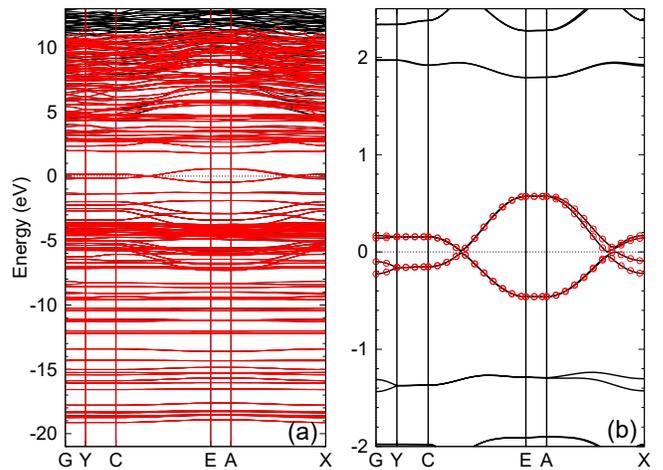}
    \caption{
     Interpolated bands, red lines in (a) and red circles in (b), of TTF-TCNQ, 
     calculated by the TB Hamiltonian derived from CWFs with
     (a) $\varepsilon_0=-22.0$ and $\varepsilon_1=1.0$ eV relative to the chemical potential, 
         $k_{\rm B}T=2.0$ eV, and hybrid atomic valence orbitals as $\{Q\}$,
     and  
     (b) $\varepsilon_0=-1.0$ and $\varepsilon_1=1.0$ eV relative to the chemical potential, 
         $k_{\rm B}T=0.01$ eV, and the 26th MO and the 37th MO for TTF and TCNQ as $\{Q\}$. 
     The solid line is the original one directly calculated by the conventional scheme. 
     The number of k-points for the Brillouin zone sampling was 5 $\times$ 19 $\times$ 3.
     An experimental crystal structure was used \cite{Kistenmacher1974}. 
     The values of the DM function per CWF are 0.441 and 0.022 for (a) and (b), respectively.
    }
   \label{fig:TTF-TCNQ-band}
\end{figure}

The interpolated bands for a molecular crystal, consisting of tetrathiafulvalene (TTF) 
and tetracyanoquinodimethane (TCNQ), are shown in Figs.~\ref{fig:TTF-TCNQ-band} (a) and (b).
By employing hybrid atomic orbitals as the guiding functions $\{Q\}$, the wide range of bands are reproduced 
as shown in Fig.~\ref{fig:TTF-TCNQ-band} (a). On the other hand, as shown in Fig.~\ref{fig:TTF-TCNQ-band} (b),
only four bands near the chemical potential are reproduced using the embedded MOs in the bulk 
as the guiding functions $\{Q\}$. 
The MOs were calculated by grouping the TTF and TCNQ molecules separately as explained 
in Sec.~III B. The expectation values calculated by Eq.~(\ref{eq:eignormalized_y}) and the singular values $\omega$ 
of $\Lambda_{g}$ are listed in Table~\ref{table:MO_data}. It can be seen that the singular values largely change 
from $\omega_{26}$ to $\omega_{27}$ and from $\omega_{37}$ to $\omega_{38}$ for TTF and TCNQ, respectively. 
So, the 26th MO and the 37th MO for TTF and TCNQ, respectively, were used as the guiding functions $\{Q\}$. 
Since there are two TTF molecules and two TCNQ molecules in the unit cell, we have the four embedded MOs as $\{Q\}$, resulting 
in the four bands. In Figs.~\ref{fig:CWFs} (c) and (d) two of the resultant CWFs are shown, which localize in 
TTF and TCNQ molecules, respectively, as expected.
The value of the DM function per CWF is found to be 0.022, which implies that the guiding MOs are very close to 
the resultant CWFs. The example demonstrates that the proposed method provides a direct way to calculate 
CWFs for targeted bands in molecular crystals. 

\begin{figure}[t]
    \centering
    \includegraphics[width=8.5cm]{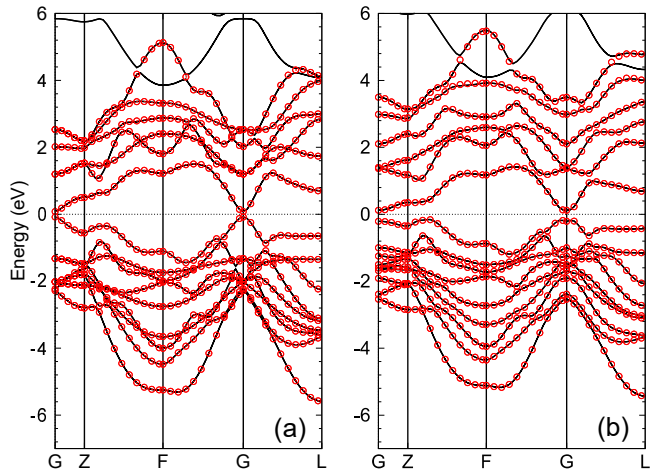}
    \caption{
     Interpolated bands (red circles) of Bi$_2$Se$_3$ (a) without and (b) with the spin-orbit interaction, 
     calculated by the TB Hamiltonian derived from CWFs.
     The solid line is the original ones directly calculated by the conventional scheme. 
     $\varepsilon_0=-6.0$ and $\varepsilon_1=3.0$ eV relative to the chemical potential, $k_{\rm B}T=1.0$ eV, 
     and hybrid Bi $6p$-orbitals and Se $4p$-orbitals as $\{Q\}$, were used in both the calculations.  
     The number of k-points for the Brillouin zone sampling was 7 $\times$ 7 $\times$ 7.
     An experimental crystal structure was used \cite{Nakajima1963}. 
     The values of the DM function per CWF are 0.163 and 0.184 for (a) and (b), respectively.
    }
   \label{fig:Bi2Te3-band}
\end{figure}

We extend the proposed method to the non-collinear DFT \cite{vonBarth1972,Kubler1988} with spin-orbit interaction (SOI) \cite{Theurich2001}. 
The KS orbitals are expressed by two-component spinor, and the SOI is introduced by fully relativistic $j$-dependent PPs \cite{OpenMX}. 
The theoretical framework is readily extended to the non-collinear DFT with the SOI without any difficulty. 
However, it should be noted that the hybrid atomic orbitals and embedded MOs in molecules and bulks as the guiding 
functions $\{Q\}$ can be complex functions in the case. 
In Figs.~\ref{fig:Bi2Te3-band} (a) and (b), we show the Wannier interpolated bands of Bi$_2$Se$_3$ which is known to
be a topological insulator when the SOI is included in the DFT calculation. 
For the cases without and with the SOI, corresponding to Figs.~\ref{fig:Bi2Te3-band} (a) and (b), respectively, 
it is confirmed that the interpolated bands accurately reproduce the original ones regardless of existence of the band inversion, 
demonstrating a wide variety of applicability of the proposed method.

\subsection{Effective Atomic Charges}

From the construction, one can calculate CWFs closest to (hybrid) valence atomic orbitals, 
while fully respecting the occupied states. The CWFs provide us a unique way to evaluate effective atomic charges.
Since the CWFs are a set of orthonormal functions, the effective atomic charge $\kappa_i$ of atom $i$ is easily calculated
without arbitrariness as 
 \begin{eqnarray}
   \nonumber
   \kappa_i
     &=&
    N_i^{\rm (val)}
    -
    \sum_{p\in {\rm atom}~i}\langle W_{{\bf 0}p}\vert \hat{P}\vert W_{{\bf 0}p}\rangle,\\
     &=&
    N_i^{\rm (val)}
    -
    \frac{1}{N_{\rm BC}}
    \sum_{p\in {\rm atom}~i}\sum_{\bf k,\mu} 
     f(\varepsilon_{\bf k \mu}) u_{{\bf k}\mu,p}^{*} u_{{\bf k}\mu,p}, 
  \label{eq:Charge}
 \end{eqnarray}
where $\hat{P}$ is the projection operator defined by Eq.~(\ref{eq:PO}), 
and $N_i^{\rm (val)}$ is the number of valence electrons in the PP of atom $i$.  
Though the summation over orbitals $p$ belonging to atom $i$ is taken into account in Eq.~(\ref{eq:Charge}), 
one can analyze the orbital resolved charges as well. 
Table~\ref{table:Charges} shows effective atomic charges in a HCN molecule and the NaCl bulk, calculated by Eq.~(\ref{eq:Charge}) 
and the Mulliken population analysis \cite{Mulliken1955}. Both the systems are known to be notorious cases 
because of the difficulty in calculating the effective charges \cite{Knizia13}.
We see that the effective charges by the CWFs quickly converge as a function 
of basis functions, while those by the Mulliken population are highly dependent on the choice of the basis function.
The effective atomic charges obtained from the CWFs could serve as a valuable tool for analyzing electronic
structures in a manner related to the approach discussed in Ref.~\cite{Knizia13}.
Further investigation in this direction will be conducted in future studies.

   \begin{table}[t]
    \caption{
     Effective atomic charges in a HCN molecule and the NaCl bulk, calculated by the proposed method (CWF) 
     and Mulliken population analysis (MP) with a variety of basis functions, where $A$ represents the constituting atoms.
     Hybrid atomic valence orbitals are used as the guiding functions $\{Q\}$. 
     For the window function, 
     $\varepsilon_0=-55.0$ and $\varepsilon_1=0.0$ eV for HCN and $\varepsilon_0=-35.0$ and $\varepsilon_1=0.0$ eV for NaCl,
     relative to the chemical potential, and $k_{\rm B}T=0.1$ eV are used. 
     The number of k-points for the Brillouin zone sampling is 9 $\times$ 9 $\times$ 9 for the NaCl bulk 
     with the lattice constant of 5.63~\AA.
    }
   \vspace{3mm}
   \begin{tabular}{ccccccccc}
   \hline\hline
     HCN & & CWF &&&&& MP\\
   \hline
     Basis  & H & C & N &&& H & C & N\\
     \hline
      $A$6.0-s1p1     & 0.077 & -0.052 & -0.025 &&& 0.384 & -0.164 & -0.221\\
      $A$6.0-s2p2     & 0.069 &  0.003 & -0.073 &&&-0.070 &  0.321 & -0.252\\
      $A$6.0-s2p2d1   & 0.066 &  0.009 & -0.076 &&&-0.008 &  0.393 & -0.385\\
      $A$6.0-s3p3d2   & 0.066 &  0.008 & -0.074 &&& 0.110 &  0.298 & -0.408\\
      $A$6.0-s3p3d2f2 & 0.066 &  0.008 & -0.074 &&& 0.167 &  0.297 & -0.464\\
   \hline
   \hline
     NaCl & CWF & &&&& MP & \\
     \hline
     Basis  & Na & Cl & &&& Na & Cl & \\
     \hline
      $A$9.0-s2p1     & 0.391 & -0.391 &  &&& 0.716 & -0.716 & \\
      $A$9.0-s3p2     & 0.422 & -0.422 &  &&& 0.595 & -0.595 & \\
      $A$9.0-s3p3d2   & 0.421 & -0.421 &  &&& 0.158 & -0.158 & \\
      $A$9.0-s3p3d2f2 & 0.421 & -0.421 &  &&&-0.062 &  0.062 & \\
  \hline
   \end{tabular}
   \label{table:Charges}
  \end{table}

\section{CONCLUSIONS}

We presented a non-iterative method to calculate the closest Wannier functions (CWFs) to a given set of localized guiding functions
such as atomic orbitals, hybrid atomic orbitals, and molecular orbitals (MOs). 
We defined the distance measure (DM) function $F$ by the sum of the squared distance between the projection functions $L$ 
and Wannier functions $W$ in the Hilbert space, and considered minimizing the DM function. 
It was shown that the minimization of the DM function is achieved by the polar decomposition of the projection matrix $A$ 
with a window function via the singular value decomposition (SVD) in a non-iterative manner. 
The CWFs can be uniquely constructed as long as the projection matrix $A$ is positive definite. 
It was also shown that the method is free from subtle choice of the gauge. 
The disentanglement of bands is naturally taken into account by introducing a smoothly varying window function, 
and including $N_{\rm band}$ Bloch functions to generate $N_{\rm CWF}$ CWFs, where $N_{\rm CWF} \le N_{\rm band}$. 
Even for isolated bands, the method always performs the disentanglement of bands without any difficulty. 
We have implemented the proposed method into the OpenMX code, which is based on the density functional theory (DFT), 
numerical pseudo-atomic orbitals (PAOs), and norm-conserving pseudopotentials (PPs), 
and introduced three types of guiding functions, i.e., atomic orbitals,  hybrid atomic orbitals, and 
embedded MOs in molecules and bulks. The first two are easily employed from the PAOs. 
For the last one, we developed a method to calculate embedded MOs in molecules and bulks, which focuses on a partial 
trace formula of the projection operator for the occupied space and applies SVD to the partial matrix for the projection operator. 
The interpolated bands by tight-binding (TB) models derived from the CWFs reproduce well the targeted conventional bands of 
a wide variety of systems including Si, Cu, the TTF-TCNQ molecular crystal, and a topological insulator of Bi$_2$Se$_3$. 
These successful reproduction of targeted bands clearly demonstrates a wide variety of applicability of the proposed method.
We further show the usefulness of the proposed method in calculating effective atomic charges, implying that  
the CWFs closest to atomic orbitals can be used as a {\it measure} to analyze electronic structures from one system to the others.
Thus, we conclude that the proposed method is an alternative way in efficiently calculating WFs, and the concept of CWFs
will provide a basis for development of novel methods of analyzing electronic structures and calculating physical properties.

\begin{acknowledgments}
The author wishes to express gratitude to Prof. Yoshiaki Sugimoto for inspiring the development of CWFs through collaborative research. 
Thanks are also extended to Mr. Ryotaro Koshoji for his insightful comments on the theoretical aspect. 
Part of the computation in this study was carried out using the computational facility of 
the Institute for Solid State Physics at the University of Tokyo.

\end{acknowledgments}

\appendix

\setcounter{figure}{0} \renewcommand{\thefigure}{A.\arabic{figure}}

\section{The upper bound of the diagonal elements of the matrix $D$}

 A proof for the upper bound of the diagonal elements of the matrix $D$ is given in the appendix. 
 Note that $UV$ and $BV$ in Eq.~(\ref{eq:MatB}) are partial unitary matrices in size of 
 $N_{\rm band}\times N_{\rm CWF}$.
 Since $D$ is hermitian, its eigenvalues are real. Let $\lambda$ be 
 an eigenvalue of $D$ and $x$ be the corresponding eigenvector. 
 Then, the following equation holds:
 \begin{eqnarray}
   \frac{1}{2}\left(V^{\dag}U^{\dag}BV+V^{\dag}B^{\dag}UV \right)\vert x \rangle = \lambda \vert x\rangle. 
   \label{eq:AP1-EVeq1}
 \end{eqnarray}
 Defining $\vert y\rangle = UV\vert x \rangle$ and $\vert z\rangle = BV\vert x \rangle$, and 
 operating $\langle x\vert$ from the left side of Eq.~(\ref{eq:AP1-EVeq1}), we have 
 \begin{eqnarray}
   \langle y\vert z \rangle + \langle z\vert y \rangle = 2\lambda \langle x\vert x\rangle. 
   \label{eq:AP1-EVeq2}
 \end{eqnarray}
 Noting $\langle y\vert z \rangle + \langle z\vert y \rangle \le 2\vert\langle y\vert z \rangle \vert$, 
 and using the Cauchy-Schwarz inequality 
 $\vert\langle y\vert z \rangle \vert\le \sqrt{\langle y\vert y \rangle \langle z\vert z \rangle}$,
 we obtain 
 \begin{eqnarray}
   \langle y\vert z \rangle + \langle z\vert y \rangle
   \le
   2\sqrt{\langle y\vert y \rangle \langle z\vert z \rangle}.
   \label{eq:AP1-ineq1}
 \end{eqnarray}
 Since $\langle y\vert y \rangle=\langle z\vert z \rangle=\langle x\vert x \rangle$, 
 combining Eq.~(\ref{eq:AP1-EVeq2}) and Eq.~(\ref{eq:AP1-ineq1}) results in 
 \begin{eqnarray}
   \lambda \langle x\vert x\rangle
   \le
    \langle x\vert x\rangle.
   \label{eq:AP1-ineq2}
 \end{eqnarray}
 Considering $\langle x\vert x\rangle\ne 0$, the upper bound of the eigenvalues of $D$ is found to be 
 \begin{eqnarray}
   \lambda \le  1. 
   \label{eq:AP1-ineq3}
 \end{eqnarray}
 Noting that the matrix $D$ can be written by $x$ and $\lambda$ as 
 \begin{eqnarray}
   D = \sum_{\nu}\vert x_{\nu} \rangle \lambda_{\nu} \langle x_{\nu}\vert, 
   \label{eq:AP1-B}
 \end{eqnarray}
 the diagonal elements $d_{nn}$ of the matrix $D$ is given by 
 \begin{eqnarray}
   d_{nn} = \sum_{\nu} \vert \langle n \vert x_{\nu} \rangle \vert^2 \lambda_{\nu},
   \label{eq:AP1-Bdiag1}
 \end{eqnarray}
 where $\langle n \vert x_{\nu}\rangle$ is the $n$-th element in the vector $x_{\nu}$. 
 Since the upper bound of $\lambda$ is unity, and $\{x\}$ forms a unitary matrix, 
 we obtain the upper bound of the diagonal elements of the matrix $D$ as   
 \begin{eqnarray}
   d_{nn} \le  1.
   \label{eq:AP1-Bdiag2}
 \end{eqnarray}

\end{document}